# Could the negative capacitance effect be used in field-effect transistors with a ferroelectric gate?

Eugene A. Eliseev[1], Anna N. Morozovska[2*], Lesya P. Yurchenko[1] and Maksym V. Strikha [3,4†]

1 - Institute for Problems of Materials Science, National Academy of Sciences of Ukraine, Krjijanovskogo 3, 03142 Kyiv, Ukraine

2 - Institute of Physics, National Academy of Sciences of Ukraine, prospect Nauky 46, 03028 Kyiv, Ukraine,

3 - Taras Shevchenko Kyiv National University, Faculty of Radiophysics, Electronics and Computer Systems, Pr. Akademika Hlushkova 4g, 03022 Kyiv, Ukraine,

4 - V.Lashkariov Institute of Semiconductor Physics, National Academy of Sciences of Ukraine, Pr. Nauky 41, 03028 Kyiv, Ukraine

**Abstract**

We analyze the distributions of electric potential and field, polarization and charge, and the differential capacitance of a silicon metal-oxide-ferroelectric field effect transistor (MOSFET), in which a gate insulator consists of thin layers of dielectric $SiO_2$ and weak ferroelectric $HfO_2$. It appeared possible to achieve a quasi-steady-state negative capacitance of the $HfO_2$ layer, $C_{HfO_2} < 0$, if the layer thickness is close to the critical thickness of the size-induced ferroelectric-paraelectric phase transition. The quasi-steady-state negative capacitance, being a very slow-varying transient state of the ferroelectric, corresponds to a positive capacitance of the whole system, and so its appearance does not break any thermodynamic principle. Implementation of the quasi-steady-state negative capacitance $C_{ins}$ of the gate insulator can open the principal possibility to reduce the MOSFET subthreshold swing below the critical value, and to decrease the gate voltage below the fundamental Boltzmann limit. However, we failed to found the parameters for which $C_{ins}$ is negative in the quasi-steady states; and thus, the negative $C_{HfO_2}$ cannot reduce the subthreshold swing below the fundamental limit. Nevertheless, the increase in $C_{ins}$, related with $C_{HfO_2} < 0$, can decrease the swing above the limit, reduce device heating during the operation cycles, and thus contribute to further improvements of the MOSFET performances.

---

* Corresponding author: anna.n.morozovska@gmail.com

† Corresponding author: maksym.strikha@gmail.com

## 1. Introduction

In recent years, there have been intense attempts to create devices that would use a negative capacitance effect [1, 2, 3] predicted by Rolf Landauer as early as 1976 [4]. If successful, such attempts would lead to a real breakthrough in the scaling of modern electronic devices [5, 6]. In particular, the use of thin ferroelectric films has been proposed for this purpose, since they can maintain ferroelectric properties at the thicknesses of 5 nm or even less; are CMOS-compatible and thermally stable in combination with silicon; allow deposition on 3D substrates; have a wider band gap than silicon and form a significant contact barrier for the electrons from the silicon conduction band. The principal possibility of reduction (in case of success) the subthreshold swing and supply voltage to the values below the fundamental limits define the great interest in the creation of these systems using ferroelectrics [7].

Thin films of a CMOS-compatible "weak" ferroelectric $HfO_2$, which spontaneous polarization, depending on the conditions, is within the range 2 – 40 $\mu C/cm^2$, band gap is equal to 5.8 eV, electron affinity to vacuum ~2 eV, and relative dielectric permittivity is about 25, are very promising candidates for metal-oxide-ferroelectric field effect transistors (**MOSFETs**) [8, 9, 10]. Below we consider a silicon MOSFET, in which the gate insulator consists of thin layers of a dielectric $SiO_2$ and a weak ferroelectric $HfO_2$. We study the possibility of implementing a stable negative capacitance of the insulator in such a system, which would open the principal possibility to reduce the subthreshold swing to the values below the threshold, 60 mV/decade at room temperature, and supply voltage to the values below the fundamental Boltzmann limit, 0.5 V, which would be an important step towards further miniaturization of MOSFETs.

## 2. Problem formulation

The scheme of the transistor in which the gate insulator is formed from thin layers of a passive dielectric $SiO_2$, a weak ferroelectric $HfO_2$, and a p-type silicon semiconducting channel is shown in **Fig. 1(a)**. **Figure 1(b)** shows the heterostructure for the case of flat bands [11, 12], when the gate voltage of flat bands (**FB**) is applied. The voltage is determined by the difference between the work-function of the gate and the semiconductor channel, as well as by the surface charge at the $HfO_2$ – Si interface. We regard that the surface charge is due to the bound charge, created by the polarization $P_3$ in the ferroelectric layer, which depends on the gate voltage, and neglect the charge on the electron traps at the interface. Thin dielectric $SiO_2$, ferroelectric $HfO_2$, and p-type semiconductor silicon layers have the thickness $d$, $h$ and $t$, respectively. Their relative dielectric permittivity are $\varepsilon_e$, $\varepsilon_b$ ("background") and $\varepsilon_s$, respectively. Corresponding electric potentials are denoted as $\phi^{(e)}$, $\phi^{(i)}$ and $\phi^{(s)}$, respectively.

The ferroelectric film is regarded either paraelectric or single-domain, that is a very probable assumption for thin HfO$_2$ films. For both cases it is characterized by an inhomogeneous one-component spontaneous polarization, $P_3(x, z)$, perpendicular to film surfaces.

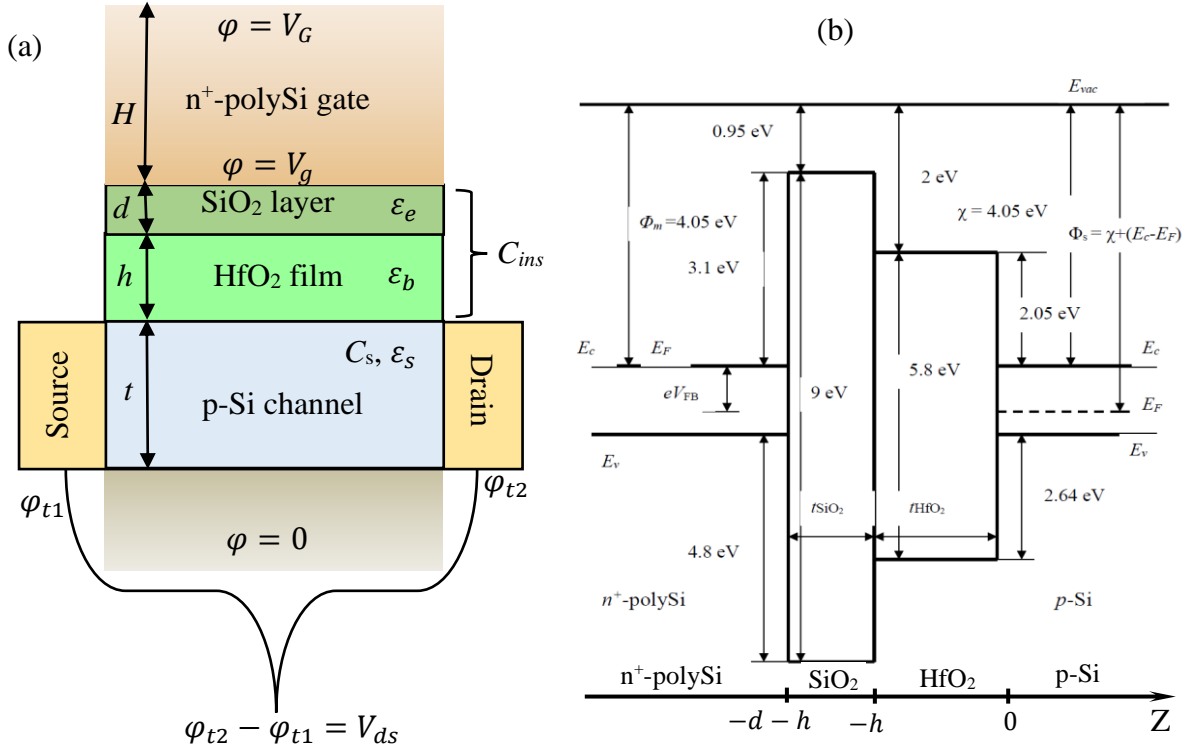

**Figure 1.** The geometry and band diagram of the considered FET, consisting of the n$^+$ poly-Si gate, dielectric SiO$_2$ layer, ferroelectric HfO$_2$ film, and the p-type Si channel.

The spatial distribution and evolution of the spontaneous polarization is described using Landau-Ginzburg-Devonshire (**LGD**) approach and electrostatic equations. The system of electrostatic governing the potential distribution have the form:

$$\varepsilon_0\varepsilon_e \left(\frac{\partial^2}{\partial z^2} + \frac{\partial^2}{\partial x^2}\right)\phi^{(e)} = 0, \qquad -h - d \leq z \leq -h, \qquad (1a)$$

$$\varepsilon_0\varepsilon_b \left(\frac{\partial^2}{\partial z^2} + \frac{\partial^2}{\partial x^2}\right)\phi^{(i)} = \frac{\partial P_3}{\partial z}, \qquad -h \leq z \leq 0, \qquad (1b)$$

$$\varepsilon_0\varepsilon_s \left(\frac{\partial^2}{\partial z^2} + \frac{\partial^2}{\partial x^2}\right)\phi^{(s)} = -e(Z_d N_d^+ - n - Z_a N_a^- + p), \qquad 0 \leq z \leq t. \qquad (1c)$$

The dielectric permittivities are $\varepsilon_e = 3.9$, $\varepsilon_b = 7$, $\varepsilon_s = 11.8$. Concentrations of the immobile ionized donors and acceptors inside the channel are $N_d^+$ and $N_a^-$, respectively; the charge densities of mobile free electrons and holes, are $n$ and $p$, respectively. When the electric current is absent the charge densities depend on the electric potential $\phi^{(s)}$ in the conventional way [13] (see **Appendix A** for details). To model the semiconducting properties of the channel we use the following parameters: acceptor concentration $N_{a0} = 10^{24}$ m$^{-3}$, their ionization degree $Z_a = 1$ and energy level $E_a = -1.05\ eV$; donor concentration $N_{d0} \ll 10^{20}$ m$^{-3}$, band gap $E_g = 1.1\ eV$. The bottom of the conduction band is selected as zero energy. The boundary conditions to Eqs.(1) are itemized below.

(a). The potential $\phi^{(e)}$ at the top of is poly-Si gate equal to the applied potential, $V_G$, $\phi^{(e)}(-H) = V_G$, where $H \gg h$. The potential $\phi^{(e)}$ at the n$^+$ poly-Si gate – SiO$_2$ interface, $z = -h - d$, is equal to the potential difference:

$$\phi^{(e)}\Big|_{z=-h-d} = V_G - V_{FB} \equiv V_g, \tag{2a}$$

where $V_{FB}$ is equal to the flat-band potential, and we regard that $V_{FB} \approx 1$eV for estimates. Below, for the sake of simplicity we recall $V_g$ as the reduced gate voltage, keeping in mind that it is differs from the potential $V_G$ applied to the top surface of the n$^+$ poly-Si gate. The introduction of $V_g$ in Eq.(2a) allows us to model three-layer structure, consisting of SiO$_2$, HfO$_2$ layers, and p-Si channel, instead of the realistic four-layer structure, shown in **Fig. 1(b).**

(b) The electric potential is continuous at the HfO$_2$ - p-Si and SiO$_2$-HfO$_2$ interfaces, $z = -h$ and $z = 0$, respectively:

$$\left(\phi^{(e)} - \phi^{(i)}\right)\Big|_{z=-h} = 0 \text{ and } \left(\phi^{(i)} - \phi^{(s)}\right)\Big|_{z=0} = 0 \tag{2b}$$

The bottom surface of the channel is grounded, $\phi^{(s)}\Big|_{z=t} = 0$. The electric current in the channel of length $L$ is controlled by the source-drain potential difference, $\phi^{(s)}(x=0) - \phi^{(s)}(x=L) = V_{ds}$.

(c) The electric displacement is also continuous at the insulator interfaces, $z = -h$ and $z = 0$:

$$\left(-\varepsilon_0\varepsilon_b \frac{\partial\phi^{(i)}}{\partial z} + P_3 + \varepsilon_0\varepsilon_e \frac{\partial\phi^{(e)}}{\partial z}\right)\Big|_{z=-h} = 0, \quad \left(-\varepsilon_0\varepsilon_b \frac{\partial\phi^{(i)}}{\partial z} + P_3 + \varepsilon_0\varepsilon_s \frac{\partial\phi^{(s)}}{\partial z}\right)\Big|_{z=0} = 0. \tag{2c}$$

(d) It should be noted that the edges of conduction and valence bands must have jumps related with the differences of corresponding Fermi levels and electron affinities at the HfO$_2$ - p-Si and SiO$_2$-HfO$_2$ interfaces respectively:

$$\chi_i - \chi_s \approx 1.05 \text{ eV}, \qquad \chi_i - \chi_s \approx -2.05 \text{ eV}. \tag{2d}$$

These conditions influence the breaks of carrier concentrations at the interfaces. In the absence of currents, these conditions follow from the conditions of local electrochemical potential (Fermi quasi-levels) continuity.

The LGD equation governing the polarization distribution has the form:

$$\Gamma\frac{\partial P_3}{\partial \tau} + \alpha P_3 + \beta P_3^3 + \gamma P_3^5 - g\frac{\partial^2 P_3}{\partial z^2} - g^*\frac{\partial^2 P_3}{\partial x^2} = -\frac{\partial\phi^{(i)}}{\partial z}, \quad -h \le z \le 0, \tag{3a}$$

Here $\Gamma$ is the Khalatnikov kinetic coefficient [14], and the corresponding Landau-Khalatnikov relaxation time $\tau_K$ of polarization reversal can be introduced as $\tau_K = \Gamma/|\alpha|$. The coefficient $\alpha = \alpha_T(T - T_C)$. The switching time $\tau_K$ can be estimated as $10^{-9} - 10^{-10}$ seconds far from the Curie temperature $T_C$. However, this estimate works for a single-domain defect-free ferroelectric HfO$_2$. Due to the different pinning mechanisms the defect-limited switching kinetics is expected to be leading to much higher, e.g., microsecond switching times. Thus, below we analyze the quasi-static polarization response, when the period of applied voltage is much higher than $\tau_K$.

LGD coefficients of HfO$_2$:Si were determined from experimental data, as described in **Appendix B**. They are: $\alpha_T = 1.72 \times 10^6$ m/(F K), $T_c = 334$ K, $\beta = 1.013 \times 10^{10}$ C$^{-4}$·m$^5$J, $\gamma = 0$, and $g = 5 \times 10^{-10}$ m$^3$/F (see for details). We vary thicknesses in the range $d$ =(5 – 10) nm, $h$ =(10 – 100) nm, and $t$=30 nm and regard that $T$=300 K.

We use the so-called natural conditions for the ferroelectric polarization,

$$\left(\frac{\partial P_3}{\partial z}\right)\bigg|_{z=-h,0} = 0, \tag{3b}$$

that support a single-domain state in the HfO$_2$ film, and neglect the carrier presence inside the dielectric and ferroelectric layers, while considering the top gate electrode as an ideal metal.

Formally, the differential capacitance of the considered heterostructure is given by expression

$$C = \frac{dQ}{dV_g}, \tag{4a}$$

where $Q$ is the charge of the top electrode. This charge, taken with the opposite sign, is equal to the sum of the bottom electrode charge $Q_b$, and total space charge in the channel $Q_s$, $Q = -Q_b - Q_s$. Assuming that the serial expression for the capacitance is valid:

$$\frac{1}{C} = \frac{1}{C_{ins}} + \frac{1}{C_s} \tag{4b}$$

where $C_{ins}$ is the differential capacitance of the insulator, consisting of the ferroelectric and dielectric layers, and $C_s$ is the differential capacitance of the channel. While the serial expression for the capacitance, $1/C_{ins} = 1/C_{SiO_2} + 1/C_{HfO_2}$, is not rigorous due to the non-local permittivity of the ferroelectric layer, which characterizes the whole system, but not its separate parts [15], we can use the expression for the upper estimate of the ferroelectric layer differential capacitance, $C_{HfO_2}$, and put $C_{SiO_2} = \frac{\varepsilon_0 \varepsilon_e}{d}$ for the dielectric layer. The channel capacitance is $C_s \cong \frac{dQ_s}{d\varphi_s}$, where $\varphi_s$ is the potential drop, and $Q_s$ is the total space charge of the channel. Roughly, the channel capacitance can be estimated as $C_s \approx \frac{\varepsilon_0 \varepsilon_s}{w}$, where $w$ is the voltage-dependent thickness of the space-charge screening layer.

From the above estimates, the differential capacitances $C_{ins}$ and $C_{HfO_2}$ are

$$\frac{1}{C_{ins}} = \frac{1}{C} - \frac{1}{C_s}, \quad C_{ins} = \frac{C}{1 - \frac{C}{C_s}}, \tag{4c}$$

$$\frac{1}{C_{HfO_2}} \approx \frac{1}{C_{ins}} - \frac{1}{C_{SiO_2}} = \frac{1}{C} - \frac{1}{C_s} - \frac{1}{C_{SiO_2}}, \quad C_{HfO_2} \approx \frac{C}{1 - \frac{C}{C_s} - \frac{C}{C_{SiO_2}}}. \tag{4d}$$

The effect of the negative ferroelectric capacitance means that $C_{HfO_2} < 0$, meanwhile the total capacitance $C$ must be positive in the thermodynamic equilibrium, because the stable negative capacitance of the system part can be achieved only consistently with positive electrostatic energy and capacity of the entire system [16]. For the same reasons $C_{ins}$ must be positive in a steady-state regime of a two-layer capacitor structure (see e.g., Ref.[17] and refs therein), while any thermodynamic limitations on its sign in the considered three-layer structure is absent, and so the situation $C_{ins} < 0$ is not excluded.

### 3. Results of the finite-element modelling for zero source-drain voltage

Typical results of the self-consistent finite-element modelling (**FEM**) of the electric potential, ferroelectric polarization, and space charge density in the considered system under the absence of source-drain voltage, $V_{sd} = 0$, are shown in **Figs. 2-5**. The figures are calculated for an ultra-thin SiO$_2$ layer with thickness $d = (1 - 2)$ nm, room temperature, $T$=300 K, variable film thicknesses $h = (5 - 50)$ nm and channel width $t = (1 - 30)$ nm [18].

The electric potential "inversion" at the dielectric/ferroelectric interface is seen in **Fig. 2(a)**, where the sign of the voltage drop in the ferroelectric film is opposite to the signs of the potential drop in the dielectric layer and in the semiconductor channel. That say, potential drops inside the dielectric and ferroelectric layers have different signs, and the local potential is inverted in some cases. Note that the inversion effect is absent at the ferroelectric-channel interface.

Exactly the inversion effect in a ferroelectric layer is treated by many authors as a manifestation of the negative capacitance effect [19, 20, 21]. At that the local electric field, as a z-derivative of the potential, is directed against the average field in the HfO$_2$, while the direction of the field in the dielectric and semiconductor coincides with the direction of the average field. When the ferroelectric film is thin enough ($h \leq 50$ nm) and the dielectric layer is enough thick ($d \geq 1$ nm), the paraelectric phase is either stable or metastable, and so the spontaneous polarization is virtually absent in the static case [see the black curve, $P_3 = 0$, at $V_g = 0$ in **Fig. 2(b)**]. The field-induced polarization, shown by colored curves in **Fig. 2(b)**, is proportional to the local electric field in the HfO$_2$ film. Under the favorable condition, the field-induced polarization contributes to the negative capacitance effect due to the negative local susceptibility. Free carriers in the channel screen the polarization bound charge and electric field; corresponding distribution of holes density is shown in **Fig. 2(c)**. For the band structure parameters, shown in **Fig. 1(b)**, the calculated electron density appeared negligibly small in comparison with the hole density in the channel. For the same reasons one can neglect the concentration of ionized donors in comparison with the concentration of ionized acceptors, shown in **Fig. 2(d)**.

The negative capacitance effect is observed in a ferroelectric layer at moderate gate voltages, $|V_g| \leq 1$ V, as shown in **Fig. 2(a)**. The negative capacitance disappears with increasing the gate voltage above 1.5 V (see **Fig. 3**). The disappearance can be explained by the influence of the ferroelectric and/or semiconductor nonlinearity, because the susceptibility of the ferroelectric layer decreases with the electric field increase; and the electric field penetration into the channel (i.e., the effective thickness $w$ of the depletion layer) depends nonlinearly on the potential difference.

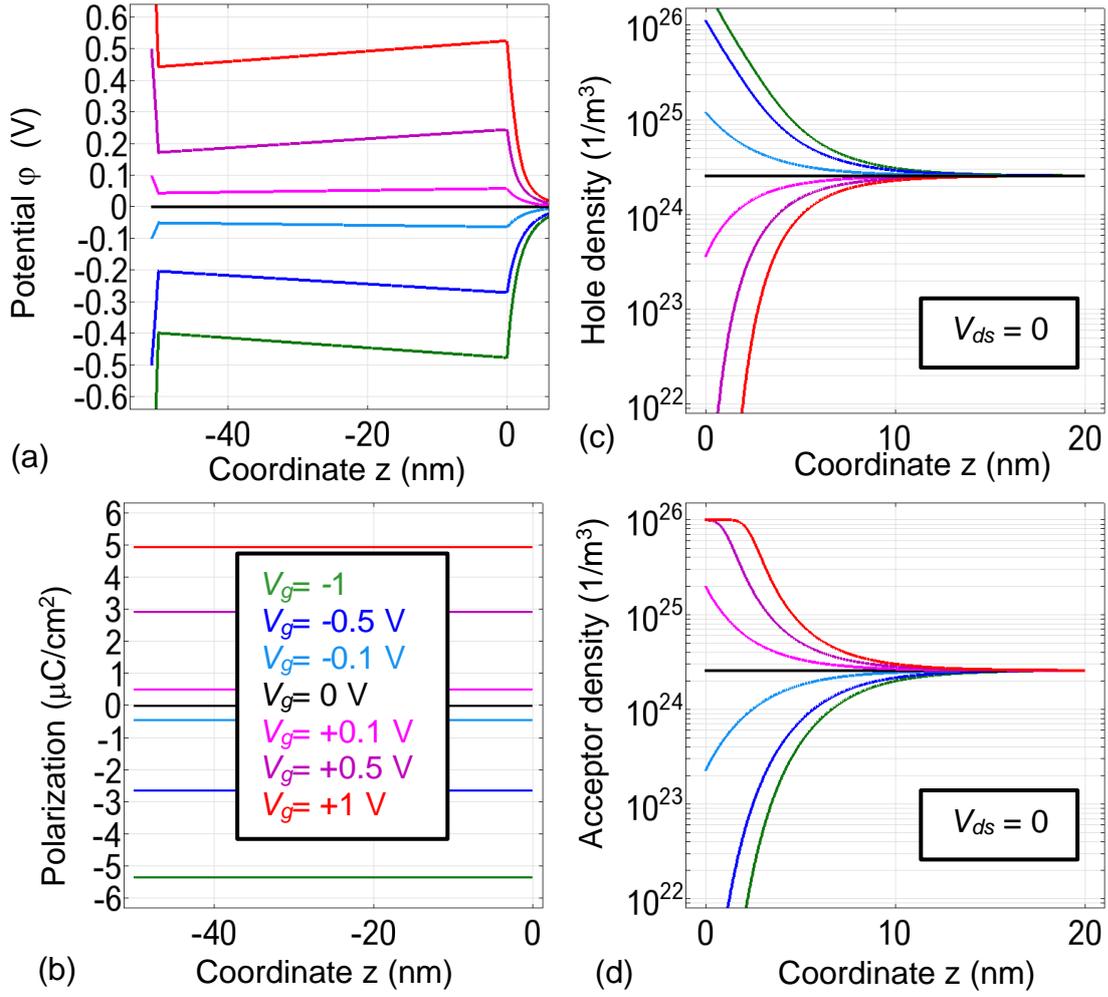

**Figure 2**. **The effect of electric potential inversion at the dielectric/ferroelectric interface.** The distribution of an electric potential in the SiO$_2$-HfO$_2$-p-Si heterostructure (**a**), electric polarization in the HfO$_2$ layer (**b**), holes density (**c**) and ionized acceptors concentration (**d**) in the p-Si channel calculated for $V_{ds} = 0$ and different reduced gate voltages: $V_g$ = -1 V (dark green curves), -0.5 V (deep blue curves), -0.1 V (blue curves), 0 (black curves), +0.1 V (magenta curves) and +1 V (red curves). The HfO$_2$ thickness $h$=50 nm and the p-Si thickness $t$=30 nm.

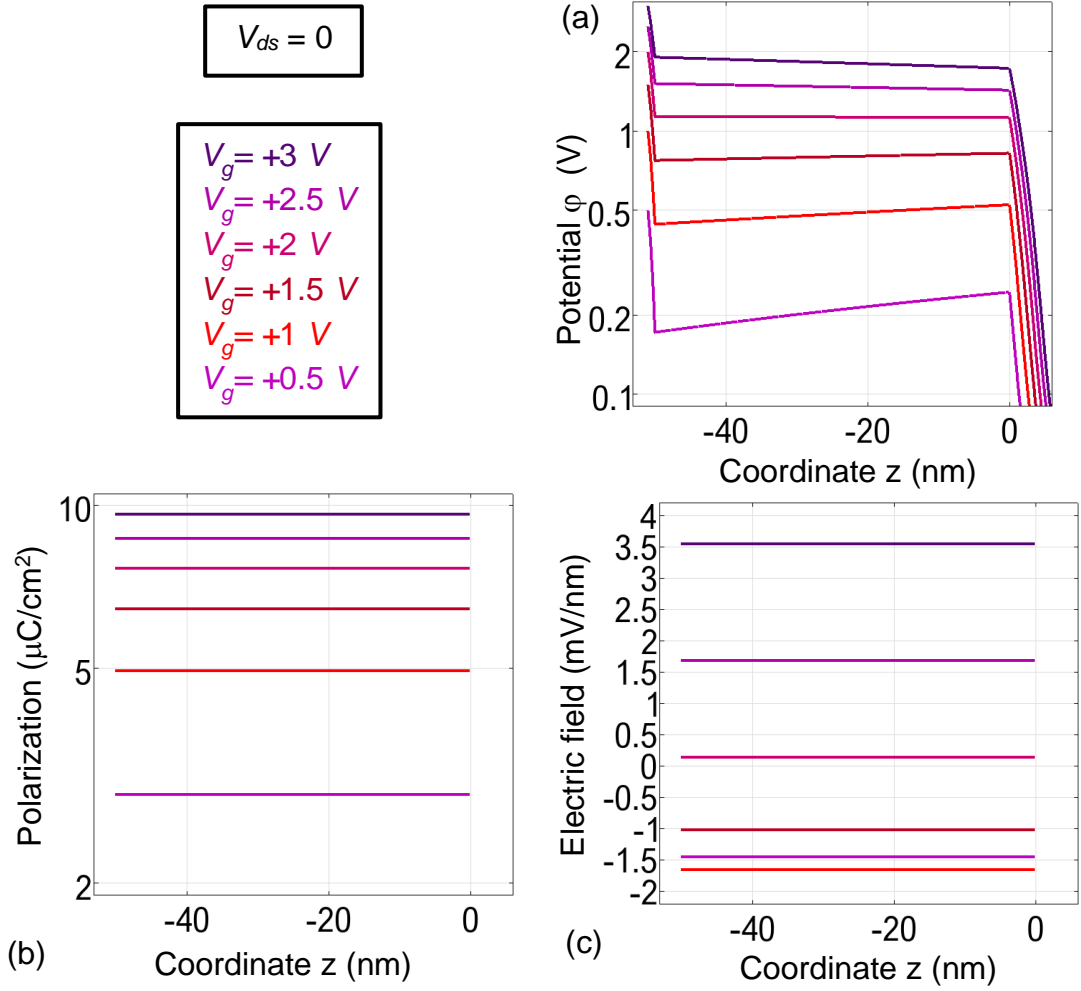

**Figure 3. Gate voltage-induced disappearance of the electric potential inversion at the dielectric/ferroelectric interface.** The distribution of an electric potential in the SiO$_2$-HfO$_2$-p-Si heterostructure **(a)**, electric polarization in the HfO$_2$ layer **(b)**, holes density **(c)** and ionized acceptors concentration **(d)** in the p-Si channel calculated for $V_{ds} = 0$ and different reduced gate voltages: $V_g = -2$ V (dark green curves), -1 V (blue curves), 0 (black curves), +1 V (magenta curves) and +2 V (red curves). Other parameters are the same as in **Fig. 2.**

Quasi-static dependences of the HfO$_2$ film polarization, $P_3$, on the gate voltage $V_g$ is shown in **Fig. 4(a)-(b)** and **Fig. 5(a)-(b)**. The dependences are calculated either for different film thicknesses $h$ [see **Fig. 4(a)-(b)**], or for different channel width $t$ [see **Fig. 5(a)-(b)**]. Polarization curves demonstrate a paraelectric-like dependence on $V_g$ for HfO$_2$ films with $h \leq 25$ nm independently on the channel widths. The height and slope of very thin ferroelectric hysteresis loops, which appear for thicker films with $h > 25$ nm (e.g., for $h = 50$ nm), depends on the channel with $t$ in a complex way. Note that the thickness for $h = 50$ nm is very close to the critical thickness of the size-induced ferroelectric-paraelectric phase transition. The loops are asymmetric with respect to the gate voltage, and this asymmetry is related with the asymmetry of the electric boundary conditions, because the top surface of HfO$_2$ film contacts with the dielectric SiO$_2$ layer, and the bottom surface contacts with the semiconductor p-Si.

Quasi-static dependences of the total differential capacitance, $C$, on the reduced gate voltage $V_g$ is shown in **Fig. 4(c)-(d)** and **Fig. 5(c)-(d)**. The dependences are calculated either for different film thicknesses $h$ [see **Fig. 4(c)-(d)**], or for different channel width $t$[see **Fig. 5(c)-(d)**]. The capacitance curves rarely demonstrate a paraelectric-like dependence on $V_g$ for considered range of channel widths (one asymmetric maxima located near $V_g = 0$). More often, the curves are ferroelectric-like with two asymmetric maximums. The asymmetry is related with the asymmetry of the polarization curves, shown in **Fig. 4(a)-(b)** and **Fig. 5(a)-(b)**, respectively. The height and separation of these maximums become more pronounced for thicker films with $h > 25$ nm and depend on the channel with $t$ in a complex way. The total capacitance is positive, and depends on $V_g$ due to the paraelectric or/and ferroelectric nonlinearity in the HfO$_2$ layer. The capacitance is maximal in the vicinity of $V_g = 0$, because the dielectric susceptibility of the HfO$_2$ layer is maximal in the voltage range, and decreases significantly with the voltage increase.

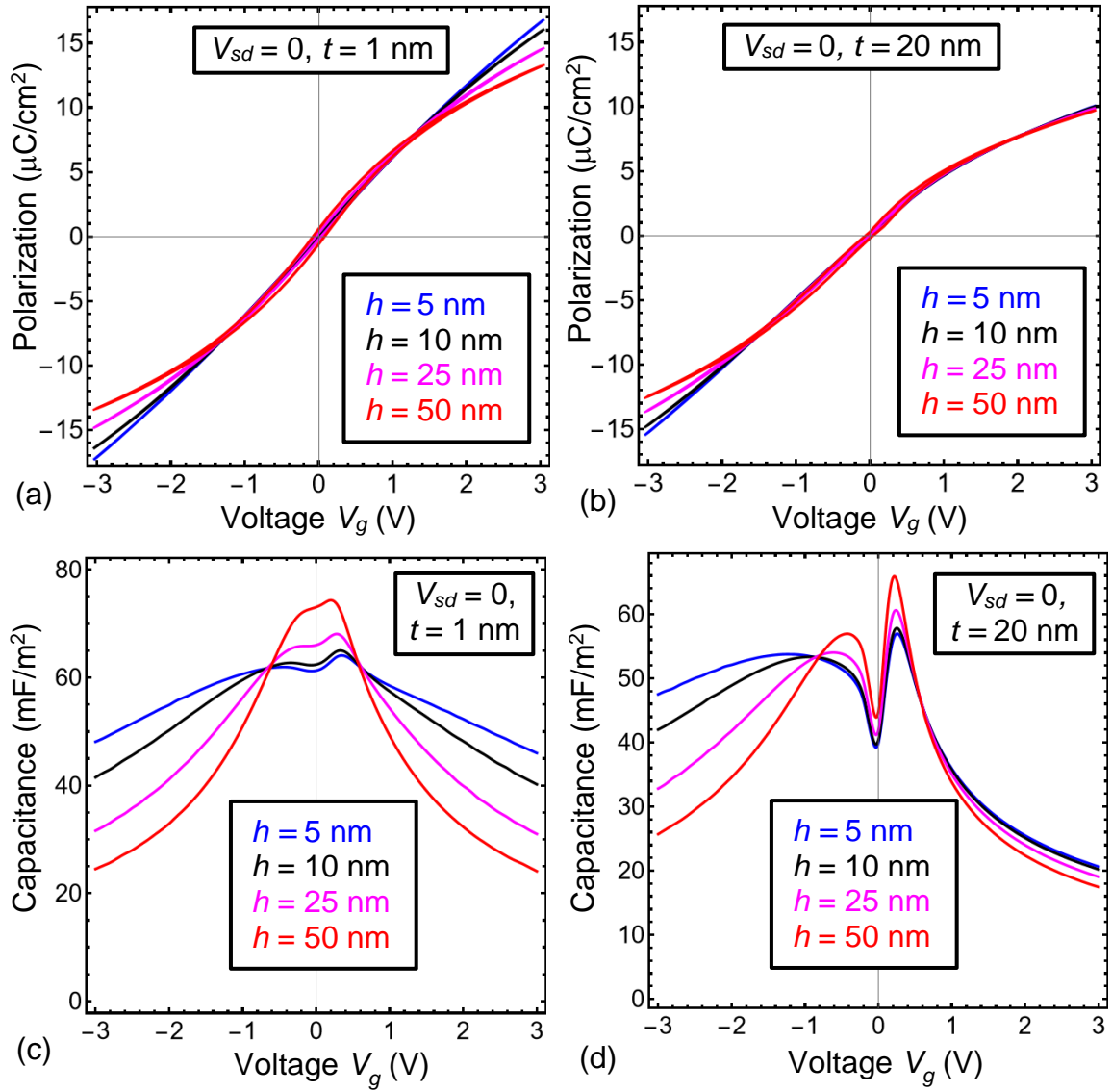

**Figure 4**. Quasi-static dependences of the HfO$_2$ film polarization (**a, b**) and total differential capacitance (**c, d**) on the reduced gate voltage $V_g$ calculated for different thickness of HfO$_2$ film $h$=5, 10, 25 and 50 nm (blue, black, magenta and red curves, respectively). The channel width $t$=1 nm in the plots (**a, c**) and $t$=20 nm in the plots (**b, d**).

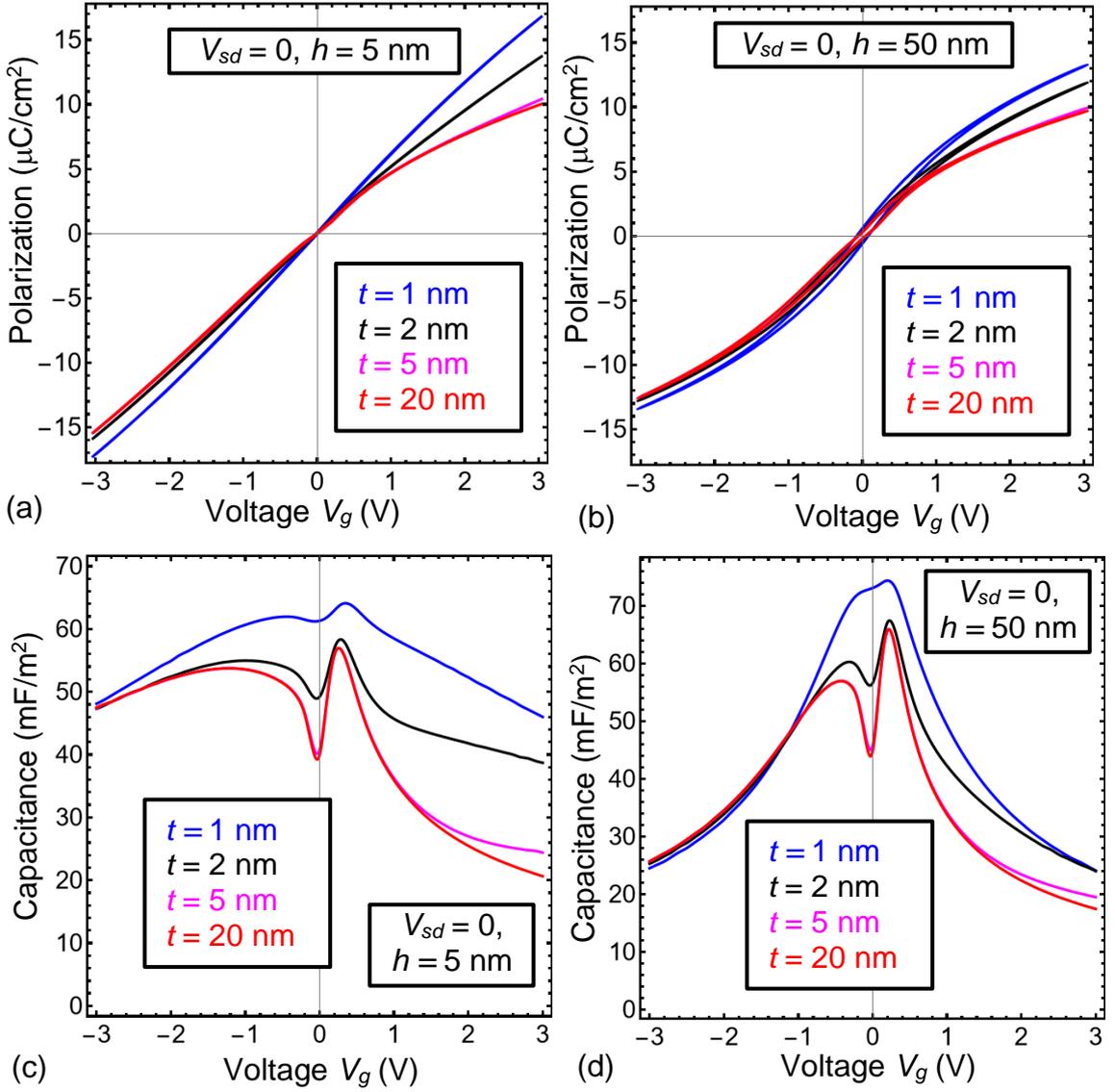

**Figure 5**. Quasi-static dependences of the HfO$_2$ film polarization **(a, b)** and total differential capacitance **(c, d)** on the reduced gate voltage $V_g$ calculated for different channel widths $t$=1, 2, 5 and 20 nm (blue, black, magenta and red curves, respectively). The HfO$_2$ film thickness $h$=5 nm in the plots **(a, c)** and $h$=50 nm in the plots **(b, d)**.

### 4. The influence of the ferroelectric capacitance the MOSFET performances

When the source-drain current flows, the electron/hole density obey continuity equations, $\frac{\partial \rho}{\partial t} + div \vec{j}_{sd} = 0$, where $\rho = e(p - n)$ is the space charge density, and $\vec{j}_{sd}$ is the density of the electric current. There are many models for the dependence of $\vec{j}_{sd}$ on the carrier densities, $p$ and $n$, and velocities, $\vec{v}_p$ and $\vec{v}_n$. In the simplest case $\vec{j}_{sd} \approx e(p\vec{v}_p - n\vec{v}_n)$, and the hole and electron velocity components satisfy Euler equations [22],

$$\frac{\partial \vec{v}_p}{\partial \tau} + \frac{\vec{v}_p}{\tau_p} + (\vec{v}_p \vec{\nabla})\vec{v}_p = -\frac{e}{m_p}\vec{\nabla}\phi^{(S)}, \qquad (5a)$$

$$\frac{\partial \vec{v}_n}{\partial \tau} + \frac{\vec{v}_n}{\tau_p} + (\vec{v}_n \vec{\nabla})\vec{v}_n = -\frac{e}{m_n}\vec{\nabla}\phi^{(S)}. \tag{5b}$$

Here $\tau_n$ and $\tau_e$ are electron and hole relaxation times, $m_n$ and $m_e$ electron and hole effective masses, respectively. It is natural to assume that $\frac{\partial v_z}{\partial z} = 0$ and $\frac{\partial v_x}{\partial z} = 0$ at the channel sidewalls, $z = 0$ and $z = t$. The boundary conditions on $\vec{v}$ at the channel terminals, $x = 0$ and $x = L$, can be natural; $\frac{\partial v_x}{\partial x} = 0$ and $\frac{\partial v_z}{\partial x} = 0$ in the quasi-stationary case, since $\frac{\partial \rho}{\partial t} = 0$ and $div\vec{j}_{sd} = 0$ in the case; and the charge density is pinned by the electrodes, $\rho = \rho_0$. For the sake of brevity, we omitted subscripts "$n$" and "$p$" in the boundary conditions for velocities.

The source-drain current $I_{ds}$, as a function of the gate and source-drain voltages, $V_g$ and $V_{ds}$, can be approximately described by a phenomenological analytical expression:

$$I_{ds} \approx \frac{V_{ds}}{R\left(1+\frac{V_{ds}}{V_s}\right)} F\left(\frac{e(V_g - V_{th})}{k_B T}\right), \tag{6a}$$

where the effective resistance $R$ and the saturation voltage $V_s$ are fitting parameters, and the function $F$ is given by expression:

$$F(\xi) = \frac{(1+\xi^2)}{(1+\xi^2)\exp(-\xi)+1} \approx \begin{cases} \exp(\xi), & \xi \ll -1, \\ \xi^2, & \xi \gg 1. \end{cases} \tag{6b}$$

The threshold voltage can be estimated as:

$$V_{th} \cong V_{FB} - \phi_{inv}^{(S)} - \frac{P_s(\phi_{inv}^{(S)})}{C_{ins}}. \tag{6c}$$

where $C_{ins}$ is the total capacitance of the ferroelectric and dielectric layers, introduced in Eq.(4b); $\phi_{inv}^{(S)}$ is a potential at the semiconductor surface, $z = 0$, corresponding the inversion of conductivity type in the channel (see e.g., [11]). The negative capacitance of a ferroelectric layer, $C_{HfO_2} < 0$, can be reached in the considered heterostructure, because the curves with inverted slope exist inside the ferroelectric layer [see **Fig. 2(a)**]. In dependence on $C_{HfO_2}$ sign, the inverse capacitances, $\frac{1}{C_{HfO_2}}$ and $\frac{1}{C_{SiO_2}}$, either sum or extract to form $\frac{1}{C_{ins}}$.

Several cases, $C_{HfO_2} < 0$, are shown by blue curves in **Fig. 6(a, c)** and **6(b, d)** for HfO₂ film of thickness $h$=5 nm and $h$=50 nm, respectively. A possible physical origin of the observed negative capacitance $C_{HfO_2}$ is the situation when the quasi-static charge at the heterostructure top and bottom electrodes increases more slowly than the ferroelectric polarization. Nevertheless $C_{HfO_2} < 0$, the capacitance $C_{ins}$, calculated for the voltage-dependent differential capacitance of the channel, $C_s \cong \frac{dQ_s}{d\varphi_s}$, is positive (see solid black, blue and magenta curves in **Fig. 6**). The negative capacitance, $C_{ins} < 0$, was calculated for the voltage-independent channel capacitance $C_s \approx \frac{\varepsilon_0 \varepsilon_s}{w}$, when the space-charge layer has a constant width $w \cong 12$ nm (see dashed black, blue and magenta curves in **Fig. 6**). This example shows that the approximation of the space-charge layer with a constant width can lead to a critical mistake in the evaluation of the channel and insulator capacitances.

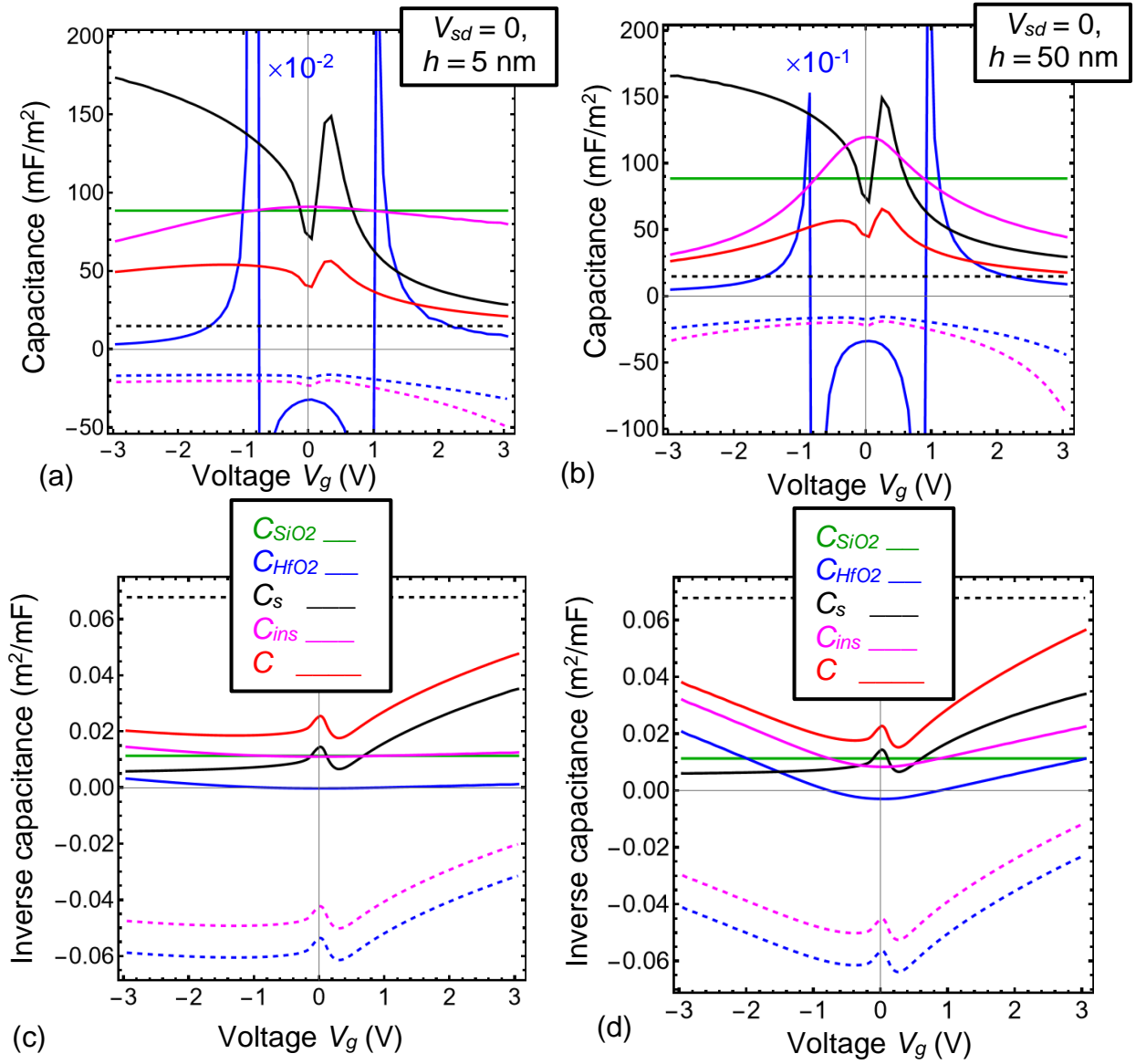

**Figure 6**. Quasi-static gate voltage dependence of the direct **(a, b)** and inverse **(c, d)** differential capacitances $C$, $C_{ins}$, $C_s$, $C_{HfO_2}$ and $C_{SiO_2}$ (solid red, magenta, black, blue and green curves, respectively) on the reduced gate voltage $V_g$. The HfO₂ film thickness $h$=5 nm in the plot **(a)** and $h$=50 nm in the plots **(b)**, the channel width $t$=20 nm. Dashed black, blue and magenta curves are calculated for the constant width of the space-charge layer, $w \cong 12$ nm.

Next, we estimate the subthreshold swing, $S$, which characterizes the gate voltage increase to achieve a tenfold increase in the drain current, as [3]:

$$S = \left[\frac{\partial(\log_{10} I_{ds})}{\partial V_g}\right]^{-1} = 2{,}3\frac{k_B T}{e}\left(1 + \frac{C_s}{C_{ins}}\right). \tag{7}$$

For positive $C_{ins}$ the minimum value of $S$, achieved at $C_s \ll C_{ins}$, is 60 mV/decade at room temperature. A possible transition to the negative capacitance of the insulator part of the heterostructure, $C_{ins} < 0$, would reduce the $S$ value below this limit, which means that the current flow in the "ON" mode will correspond to a lower source-drain voltage $V_{ds}$. This would reduce the voltage below the fundamental Boltzmann limit of 0.5 V, which would open the possibility of further miniaturization of the MOSFET, currently operating at threshold voltages not less than 0.7 V.

The capacitance $C_{ins}$ must be positive in a steady-state regime, while any thermodynamic limitations on the capacitance $C_{ins}$ of transient states are absent. However, we failed to found the conditions for which $C_{ins} < 0$ either in the steady-state or in the quasi-steady-state (see e.g., the solid magenta curves in **Fig. 6**). Instead, we obtained the increase of $C_{ins}$ and the decrease of $C_s$ under the simultaneous validity of the conditions $C_{HfO_2} < 0$ and $C_{SiO_2} > 0$ in the vicinity of $V_g \approx 0$ [see the wide maximum at the violet curves and the sharp minima at the black solid curves at $V_g \approx 0$ in **Fig. 6**]. Since $C_{ins}$ is maximal and $C_s$ is minimal in the vicinity of $V_g \approx 0$, the ratio $\frac{C_s}{C_{ins}}$ is minimal in the vicinity of $V_g \approx 0$, and the condition $0 < \frac{C_s}{C_{ins}} \ll 1$, which decreases $S$, can be valid at small $V_g$.

Since the capacitance $C_{ins}$, but not the negative capacitance of the ferroelectric layer itself, determines the FET working performances, and, in particular, the subthreshold swing $S$, the steady-state negative capacitance of a ferroelectric layer cannot reduce $S$ below the fundamental limit. Nevertheless, the increase of $C_{ins}$ can decrease $S$ above the fundamental limit in the case $\frac{C_s}{C_{ins}} \ll 1$; also, it can significantly reduce the transient losses and MOSFETs heating during the operation cycles.

## 5. Conclusions

We analyze the distributions of electric potential and field, polarization and charge, and the differential capacitance of a silicon MOSFET, in which a gate insulator consists of thin layers of dielectric $SiO_2$ and weak ferroelectric $HfO_2$.

It appeared possible to achieve a quasi-steady-state negative capacitance of a ferroelectric layer if its thickness is close to the critical thickness of the size-induced ferroelectric-paraelectric phase transition. The quasi-steady-state negative capacitance of the ferroelectric, that is very slow-varying transient state, corresponds to a positive capacitance of the whole system, and so it does not break any thermodynamic principle.

Since the capacitance of the gate insulator, $C_{ins}$, but not the negative capacitance of the ferroelectric layer itself, determines the FET working performances, implementation of the quasi-steady-state negative capacitance $C_{ins}$ can open the principal possibility to reduce the subthreshold swing $S$ below the critical value, and to decrease the gate voltage below the fundamental Boltzmann limit. However, our general conclusion is that $C_{ins} > 0$ in the quasi-steady states, meaning that the negative capacitance does not penetrate "outside" the ferroelectric layer of MOSFET, and so it cannot reduce subthreshold swing below the fundamental limit. This conclusion agrees with Ref.[16], where it was obtained in the framework of general thermodynamic principles.

Instead, we obtained the increase of $C_{ins}$ under the simultaneous validity of the conditions $C_{HfO_2} < 0$ and $C_{SiO_2} > 0$ at small gate voltages. The increase of $C_{ins}$ can decrease $S$ above the fundamental limit; also, it can significantly reduce the transient losses and MOSFETs heating during the operation cycles.

**Data availability statement.** The calculations were performed and visualized in Mathematica 12.2 ®Wolfram Research software (https://notebookarchive.org/2021-12-cwwdzxu).

**Acknowledgements.** Authors are grateful to Dr. Pavlo Zubko for very useful explanations and discussions. A.N.M. acknowledges the National Academy of Sciences of Ukraine (program 1230) and the Research and Innovation Staff Exchanges (RISE) project №778070, "Transition metal oxides with metastable phases: a way towards superior ferroic properties". M.V.S. acknowledges Taras Shevchenko National University of Kyiv for partial support of this work.

**Authors' contribution.** A.N.M. and M.V.S. generated the research idea, and A.N.M. formulated the problem mathematically. E.A.E. and L.P.Y. performed FEM. A.N.M. and M.V.S. wrote the manuscript. All co-authors discussed the results.

**Appendix A**

**A1. The general problem statement for FEM**

Let us consider a thin ferroelectric film between a passive dielectric layer and a semiconductor layer (channel). The different layers have thickness $d$, $h$ and $t$; while dielectric permittivity values are $\varepsilon_e$, $\varepsilon_b$ ("background") and $\varepsilon_s$ respectively for dielectric, ferroelectrics and semiconductor regions. Their electrostatic potentials are denoted as $\phi^{(e)}$, $\phi^{(i)}$ and $\phi^{(s)}$ respectively. The ferroelectric is characterized by an inhomogeneous spontaneous polarization with the one component $P_3$ perpendicular to the film surfaces. Neglecting the electric currents across the system one could easily wrote the system of equations governing the polarization and potential distribution using the minimization of the free energy with respect to polarization and potential, namely:

$$\varepsilon_0 \varepsilon_e \left(\frac{\partial^2}{\partial x^2} + \frac{\partial^2}{\partial y^2} + \frac{\partial^2}{\partial z^2}\right) \phi^{(e)} = 0. \quad \text{at} \quad z \in [-h-d, -h] \quad \text{(A.1c)}$$

$$\varepsilon_0 \varepsilon_b \left(\frac{\partial^2}{\partial x^2} + \frac{\partial^2}{\partial y^2} + \frac{\partial^2}{\partial z^2}\right) \phi^{(i)} = \frac{\partial P_3}{\partial z}, \quad \text{at} \quad z \in [-h, 0] \quad \text{(A.1b)}$$

$$\alpha P_3 + \beta P_3^3 + \gamma P_3^5 - g_{11} \frac{\partial^2 P_3}{\partial z^2} - g_{44} \left(\frac{\partial^2 P_3}{\partial x^2} + \frac{\partial^2 P_3}{\partial y^2}\right) = -\frac{\partial \phi^{(i)}}{\partial z}, \quad \text{at} \quad z \in [-h, 0] \quad \text{(A.1a)}$$

$$\varepsilon_0 \varepsilon_s \left(\frac{\partial^2}{\partial x^2} + \frac{\partial^2}{\partial y^2} + \frac{\partial^2}{\partial z^2}\right) \phi^{(s)} = -e(Z_d N_d^+ - n - Z_a N_a^- + p) . \quad \text{at } z \in [0, t] \quad \text{(A.1c)}$$

Here we introduced the density of the different types of charges inside the semiconductor, depending on the local potential in the following way [23]:

$$N_d^+(\phi) = N_d^0 \, f\left(\frac{-E_d + eZ_d\phi + E_f}{k_B T}\right), \tag{A.2a}$$

$$n(\phi) = N_C F_{1/2}\left(\frac{e\phi + E_f - E_C}{k_B T}\right), \tag{A.2b}$$

$$N_a^-(\phi) = N_a^0 f\left(\frac{E_a - eZ_d\phi - E_f}{k_B T}\right), \tag{A.2c}$$

$$p(\phi) = N_C F_{1/2}\left(\frac{-e\phi - E_f + E_V}{k_B T}\right). \tag{A.2d}$$

for donor, electrons, holes and acceptors respectively. Here $E_d$ and $E_a$ are the donors and acceptors energy levels respectively, $T$ is the absolute temperature, $k_B$ is a Boltzmann constant, $\varepsilon_0$ is the universal electric constant; $E_C$ is the position of conduction band bottom, while $E_V$ is the position of valence band top; effective density of states in the conductive band $N_C = \left(\frac{m_n k_B T}{2\pi \hbar^2}\right)^{3/2}$, $m_n$ is the effective mass of electrons. In Eq.(A.2d) we supposed that the effective masses of holes and electrons are the same.

In (A.2) we introduced the Fermi-Dirac distribution function and the Fermi integral

$$f(\xi) = \frac{1}{1 + \exp(\xi)} \tag{A.3a}$$

$$F_{1/2}(\xi) = \frac{2}{\sqrt{\pi}} \int_0^\infty \frac{\sqrt{\zeta} d\zeta}{1 + exp(\zeta - \xi)} \tag{A.3b}$$

$E_F$ is the Fermi energy level in equilibrium, which is determined from the electroneutrality condition $Z_d N_d^+ - n - Z_d N_a^- + p \equiv 0$ at zero potential, namely:

$$Z_d N_d^0 f\left(\frac{E_F - E_d}{k_B T}\right) + N_C F_{1/2}\left(\frac{E_f - E_C}{k_B T}\right) = N_C F_{1/2}\left(\frac{E_F - E_C}{k_B T}\right) + N_a^0\left(\frac{E_a - E_f}{k_B T}\right) \tag{A.4}$$

The potential $\phi^{(e)}$ at the n$^+$poly-Si gate – SiO$_2$ interface is equal to

$$\phi^{(e)}\big|_{z=-h-d} = V_G - V_{FB}, \tag{A.5a}$$

where $V_{FB} \approx 1\text{eV}$ is equal to the flat-band potential. The electric potential at the SiO$_2$-HfO$_2$ interface has no jump:

$$\left(\phi^{(e)} - \phi^{(i)}\right)\big|_{z=-h} = 0, \tag{A.5b}$$

and the electric displacement is continuous at the interface

$$\left(-\varepsilon_0 \varepsilon_b \frac{\partial \phi^{(i)}}{\partial z} + P_3 + \varepsilon_0 \varepsilon_e \frac{\partial \phi^{(e)}}{\partial z}\right)\bigg|_{z=-h} = 0., \tag{A.5c}$$

The continuity condition at the HfO$_2$ - p-Si interface for potential

$$\left(\phi^{(i)} - \phi^{(s)}\right)\big|_{z=0} = 0, \tag{A.5d}$$

and electric displacements

$$\left( -\varepsilon_0 \varepsilon_b \frac{\partial \phi^{(i)}}{\partial z} + P_3 + \varepsilon_0 \varepsilon_s \frac{\partial \phi^{(s)}}{\partial z} \right) \bigg|_{z=0} = 0. \tag{A.5e}$$

The bottom surface of the channel is grounded, $\phi^{(s)}|_{z=t} = 0$. The electric current in the channel of length $L$ is controlled by the source-drain potential difference, $\phi^{(s)}(x=0) - \phi^{(s)}(x=L) = V_{ds}$.

Also, we use the so-called natural conditions for ferroelectric polarization, $\left( \frac{\partial P_3}{\partial z} \right) \big|_{z=-h,0} = 0$, that support the single-domain state in HfO$_2$, and neglect the carriers' presence inside the dielectric and ferroelectric layers, while considering top gate electrode as an ideal metal.

The following approximation for the Fermi integral (A.3b) can be used to simplify numerical calculations:

$$F_{1/2}(\xi) \approx \frac{1}{\exp(-\xi) + \frac{3\sqrt{\pi}}{4}(4+\xi^2)^{-\frac{3}{4}}}. \tag{A.6a}$$

It is based on the two limiting cases when the Fermi integral could be reduced to elementary functions:

$$F_{1/2}(\xi)|_{\xi \to \pm \infty} \to \begin{cases} \exp(\xi) & \text{at } \xi \to -\infty \\ \frac{4\,\xi^{3/2}}{3\sqrt{\pi}} & \text{at } \xi \to +\infty \end{cases} \tag{A.6b}$$

the electric field, $\vec{E} = -\nabla \varphi$ could be calculated from the electrostatic potential $\varphi$ determined from the equation above.

### A2. Boundary conditions at the hetero-interface

According to Anderson [24] the electrostatic potential, vacuum level, normal components of electrons and holes currents as well as corresponding (pseudo-) Fermi levels are continuous at the interface between dissimilar semiconductors. However, conduction-band edges and valence-band edges may have breaks at the interfaces, which magnitude is determined by the affinity values as well as difference in Fermi levels in corresponding semiconductors

Therefore, the following interface conditions of continuity should be imposed for the following values:

1). Electrostatic potential;

2). Fermi pseudo-levels for holes and electrons;

3). Normal components of currents of holes and electrons;

4). Normal components of electric displacement.

Introducing Fermi pseudo-levels and their relations with electrons and holes concentrations

$$E_{nf} = E_C - e\phi + k_B T\, F_{1/2}^{-1}\left(\frac{n}{N_C}\right), \tag{A.7a}$$

$$E_{pf} = E_V - e\phi - k_B T\, F_{1/2}^{-1}\left(\frac{p}{N_C}\right). \tag{A.7a}$$

One could write the second condition via concentrations as follows:

$$E_{nf}^- = E_{nf}^+ \Rightarrow E_C^- + k_B T\, F_{1/2}^{-1}\left(\frac{n^-}{N_C}\right) = E_C^+ + k_B T\, F_{1/2}^{-1}\left(\frac{n^+}{N_C}\right) \tag{A.8a}$$

$$E_{pf}^- = E_{pf}^+ \Rightarrow E_V^- - k_B T\, F_{1/2}^{-1}\left(\frac{p^-}{N_c}\right) = E_V^+ - k_B T\, F_{1/2}^{-1}\left(\frac{p^+}{N_c}\right) \quad (A.8b)$$

Here the superscripts "-" and "+" denote physical quantities at the left and right side of the interface correspondingly. It is seen that interfacial difference of conductance and valence bands edges, $E_C^- - E_C^+$ and $E_V^- - E_V^+$ respectively, determine the break of electrons and holes at the interface. The latter differences are determined by with the differences of electron affinities as well as positions of the Fermi at equilibrium (without currents and bias).

**Appendix B. Landau model for the temperature dependence of polarization of thin HfO$_2$:Si films**

In order to fit the ferroelectric polarization $P$ dependence on temperature $T$ we will use simplest possible Landau equation of state:

$$\alpha_T(T - T_c)P + \beta P^3 = E_{internal}. \quad (B.1)$$

Here $T_c$ is the Curie temperature, $E_{internal}$ is the built-in electric field, while $\beta$ is the dielectric nonlinearity coefficient.

The value of coefficient $\alpha_T$ was estimated by Lomenzo et al. [25], while the rest of the parameters could be extracted from the experimentally measured by Boscke et al. [26] polarization temperature dependence (see **Fig. B1**).

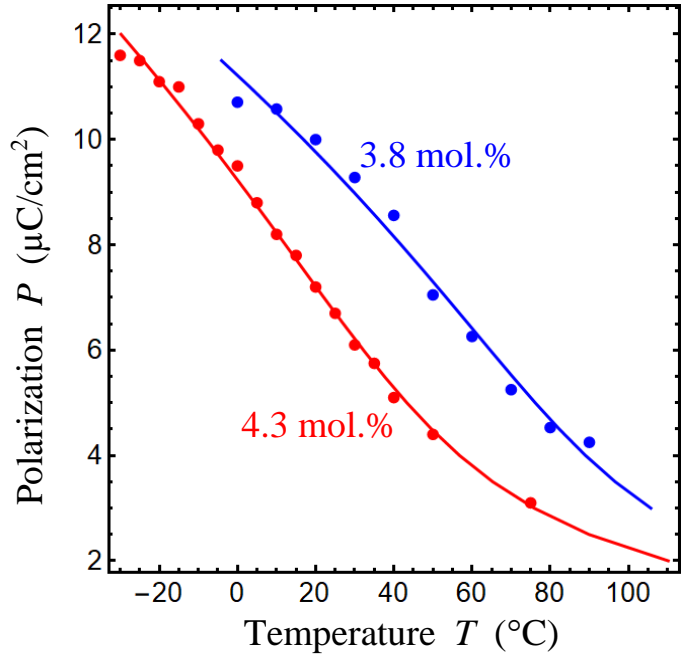

**Figure B1**. Temperature dependence of the polarization for the different degree of Si doping, namely 3.8 and 4.3 molar % (blue and red curves/symbols respectively). Points represent experimental results from [26], while curves represent the fitting with Eq.(A1). LGD parameters are summarized in Table I.

Table I. Landau-Ginsburg-Devonshire parameters of HfO$_2$ doped with Si

| Si content | $\alpha_T$ ($m\,K^{-1}F^{-1}$) | $T_c$ ($K$) | $\beta$ ($V\,m^5 C^{-3}$) | $E_{internal}$ (V/m) |
|---|---|---|---|---|
| 3.8 molar % | $1.72 \times 10^6$ | 333.9 | $1.013 \times 10^{10}$ | $2.60 \times 10^6$ |
| 4.3 molar % | $1.72 \times 10^6$ | 288.5 | $0.732 \times 10^{10}$ | $3.32 \times 10^6$ |